\documentclass[twocolumn,showpacs,preprintnumbers,amsmath,amssymb]{revtex4}

\usepackage{float}
\usepackage{graphicx}
\usepackage{dcolumn}
\usepackage{bm}

\begin{document}
\newcommand{\be}{\begin{equation}}   \newcommand{\ee}{\end{equation}}
\newcommand{\mean}[1]{\left\langle #1 \right\rangle}
\newcommand{\abs}[1]{\left| #1 \right|}
\newcommand{\set}[1]{\left\{ #1 \right\}}
\newcommand{\la}{\langle}
\newcommand{\ra}{\rangle}
\newcommand{\lb}{\left(}
\newcommand{\rb}{\right)}
\newcommand{\norm}[1]{\left\|#1\right\|}
\newcommand{\RA}{\rightarrow}
\newcommand{\tet}{\vartheta}
\newcommand{\eps}{\varepsilon}
\newcommand{\tNN}{\tilde{\mathbf{X}}_n^{NN}}
\newcommand{\NN}{\mathbf{X}_n^{NN}}
\newcommand{\ber}{\begin{eqnarray}}
\newcommand{\eer}{\end{eqnarray}}

\title{Entropy and Optimization of Portfolios}

\author{Krzysztof Urbanowicz}
 \email{krzysztof.urbanowicz@quant-technology.com}
 \affiliation{Quant Technology Sp z o.o.\\
 Modlinska 175A\\ PL--03-186 Warsaw, Poland}

\date{\today}

 \begin{abstract}
\par We briefly review the approach to optimization of portfolios according to the theory of Markowitz and propose a further modification that can improve the outcome of the optimization process. The modification takes account of the entropic contribution from the time series used to compute the parameters in the Markowitz method.
\end{abstract}
 \maketitle

\section{Background}
    \par
\begin{figure}
\centering
\includegraphics[width=8cm,angle=0]{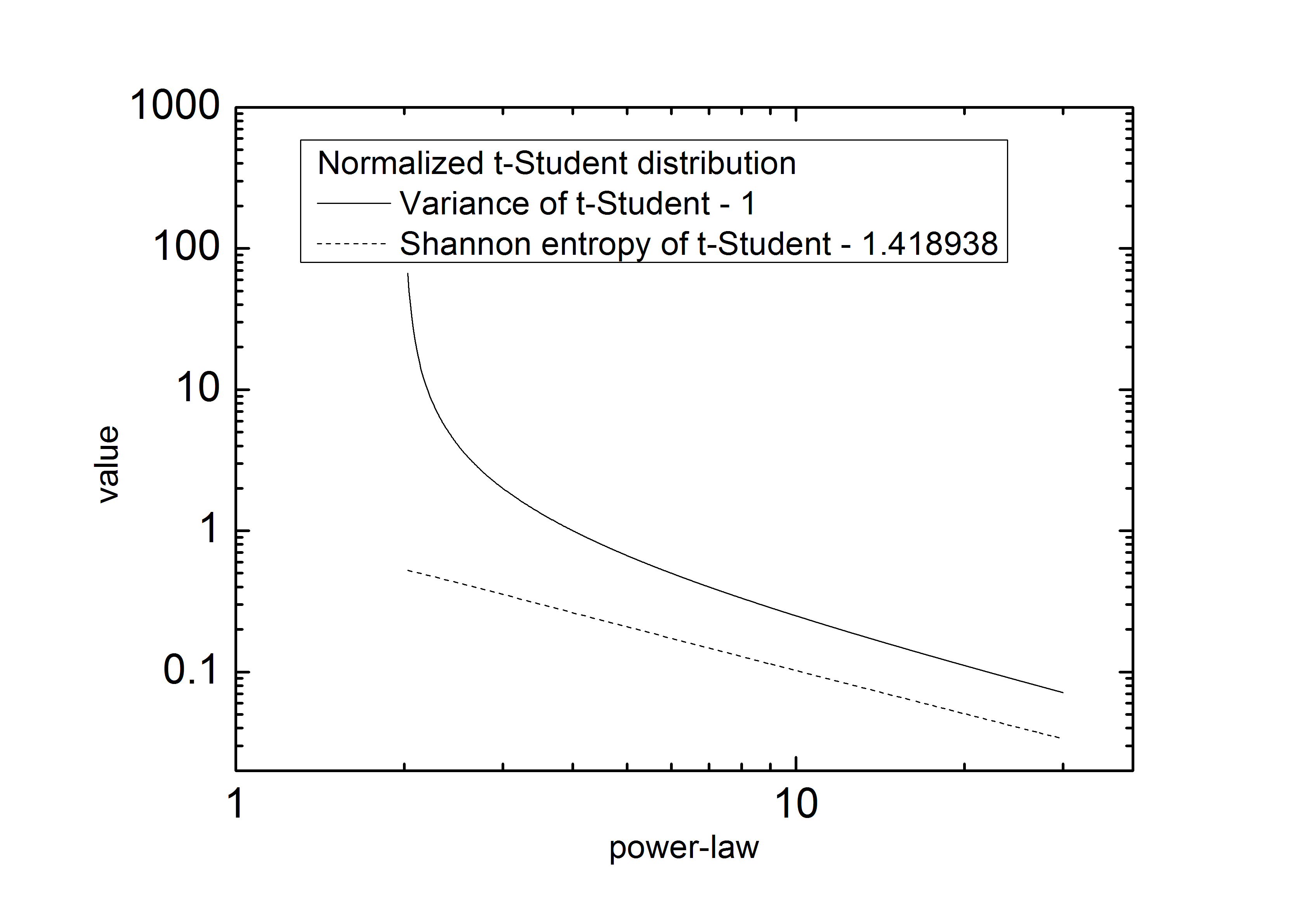}
\caption{The plot of variance and Shannon entropy of normalized t-Student distribution. The values were decreased by minimal value ie. variance and Shannon entropy of Gauss distribution, equal adequately 1 and 1.418938. \label{varianceVsEntropy} }
\end{figure}

 The theory of Markowitz and its application to optimization of stock portfolios is well documented \cite{refMarkowitz,refSiman,refTze}. The Markowitz theory shows the optimization procedure for portfolios consist from risky assets. It is assumed that we take into account past returns from each asset as well as variances and, more informational, correlations between returns of assets. This theory shows optimal way of using first and second order linear statistics to combine best portfolio. The bias of it is use of non-predictive information. We are basing on historical data. The question is if it is optimal. For sure when no additional information is available than one have to admit that 'yes'. On the other hand basing on historical returns we can calculate additional statistics like entropies \cite{refRenyi} (here we use Shannon entropy \cite{refShannon}). New information is added to the optimization from entropy of time series. Shannon entropy from probabilities of historical returns possess similar characteristic to variance of returns (look at the Fig.~\ref{varianceVsEntropy}), but not the same. One can follow difference by imagine bimodal distribution with very sharp peaks. Entropy in such case is very low (probabilities are mostly concentrated in peaks), but variance can be huge or small depends on distance between peaks. In the case of bimodal distribution Markowitz optimization will show different result depends on distance between peaks, using entropy no such situation will happen. Now, how in that situation should be an optimization. If we have two peaks of probabilities forecasting is rather simple and that is showing the entropy - the error of forecasting. In practice we often use some prediction and in that case including entropy in optimization will be more effective than only variance (in Markowitz using the covariance matrix). Concluding entropy add some new information to optimization and that is why we try to show how one should include entropy to extend Markowitz theory.
    \par Let's try to show literature background on optimization. On beginning Markowitz theory. He is assuming $M$ risky assets in portfolio which returns variate over the time interval $T$. He additionally assume, but not explicit, Gaussian distribution of returns. The average return and variance over the time interval $T$, are $m_i$ and $D_i$ respectively The overall return of the portfolio $m_p$ is then
\be m_p = \sum_{i=1}^M p_i m_i\label{eq1}\ee
The portfolio risk $D_p$ is
\be D_p = \sum_{i=1}^M p_i^2 D_i \ee
Where {$p_i$} is the set of normalized weights associated with the $M$ assets. In its simplest form the optimum weights are obtained by minimizing the utility function (minimizing the risk with established expected payoff):
\be E_p = D_p - \lambda m_p + \gamma\sum_{i=1}^M p_i \ee
The least risky portfolio, which not always are profitable in past, corresponds to that obtained by setting $\lambda=0$ leading to the optimum weights:
\be p_i^{*}=\frac{1}{ZD_i}\; where\; Z=\sum_{j=1}^M\frac{1}{D_j}\ee
A later version of the theory takes account of correlations between assets via the symmetric correlation matrix, $C_{ij}$. The risk is now defined to be:
\be D_p = \sum_{i=1}^M p_i p_j C_{ij} \label{eq5}\ee
Minimizing the risk given by definition (\ref{eq5}) and again setting $\lambda=0$ yields the optimum weights $p_i^{*}$. The least risky portfolio have no constrains on expected payoff. This is simplification of full formula:
\be p_i^{*} = \frac{1}{Z}\sum_{j=1}^M C_{ij}^{-1} \; where \; Z = \sum_{i,j=1}^M C_{ij}^{-1} \ee
As Bouchaud and Potters have pointed out, the main lesson of the theory of Markowitz' theory is the need to diversify portfolios effectively \cite{refBouchaund}. Bouchaud, Potters and Aguilar \cite{refPotter} have noted that one problem associated with the approach is that the resulting portfolio can be concentrated on only a few assets. To overcome this, they proposed including an additional constraint, the entropy of weights of assets:
\be Y_q = \sum_{i=1}^M \lb p_i^{*}\rb^q \label{eq7}\ee
The parameter $q$ was chosen to be equal to $2$ when it is seen that the term also represents the average weight of an asset in the portfolio. In general the term is, to use the language of physics, an entropic contribution to the minimization process. Indeed as Bouchaud and Potter point out, the equation (\ref{eq7}) is linearly related to the Tsallis \cite{refTsallis,refTsallis2} entropy function and the entire process of obtaining the weights, $p_i$ is equivalent to minimizing a free - utility function:
\be F_q = E - \nu \frac{Y_q-1}{q-1}\ee
Important hint we should mention. In this paper we are managing with including entropy of returns from each assets. In the paper \cite{refPotter} authors are refering to entropy of weights of assets in portfolio. This is two different entropies. Even that results are similar, we are approaching to the problem from different directions. Bouchaud, Potters and Aguilar are adding new constrain to Markowitz formula artificially, to increase diversification in practise, when errors of covariance matrix and average returns are in practical calculations with large errors. We, on the other hand, are adding entropy of time series to increase used information in optimization and it is rather natural way of such addition.
\par Bouchaud has noted another issue linked to the use of correlation matrices. This is linked to the use of finite time series when computing the $M(M-1)/2$ individual elements of the correlation matrix. If the number of assets in the portfolio becomes large it is possible that the number of data points used to compute the elements of the correlation matrix is of the same order of magnitude as the number of entries. As an example, for portfolios of the size of the $S\& P500$ where the correlation matrix contains $500$x$499/2= 124,750$ different entries. Using time series extending over two years of daily data the number of data points is $500$x$500=250,000$ which is only a factor of two larger that the number of correlation coefficients. So the statistical precision on these coefficients is subject to a large degree of measurement noise. Here we consider smaller portfolios of the order of $27$ stocks where this issue does not arise.

\par We discuss in this note another issue that can be resolved by a further modification of the theory and that leads to novel route to measure risk and further improvements in the optimization process.
\section{Approach}
\par The above discussion leads directly to the following 'free-utility' function \cite{refPotter}:
\be F=\sum_{i,j=1}^M p_i p_j C_{ij} + \alpha \sum_{i=1}^M \lb p_i \rb^2+\beta \sum_{i=1}^M p_i m_i + \gamma \sum_{i=1} ^M p_i\label{eq9}\ee
Where $\alpha$, $\beta$ and $\gamma$ are Lagrange multipliers. We now recognize that the entropy is in fact similar to the variance in that it is a measure of the level of variability in the time series and propose that it serve as an estimator of risk (this concept was noted in the first section). The entropy can be replaced by the variance only in the case of Gaussian distributions. The 'fat' tailed distributions are not fully described by a variance (see Fig.~\ref{varianceVsEntropy}) and in such a case we need more parameters. Risk related to 'fat' tailed distributions is larger than the variance so for risk evaluation we introduce a second risk factor that describes the level of possible realizations of the system in the case of a fully random process, which is measured by entropy.
\par The parameter that balances the impact of variance and entropy in physics is the temperature. Here it is the Lagrange multiplier corresponding to the entropy.
\par On the basis of this reasoning, we now propose a new functional of the weights, {$p_i$}:
\be F=\sum_{i,j=1}^M p_i p_j C_{ij} + \alpha \sum_{i=1}^M S_i \lb p_i \rb^2+\beta \sum_{i=1}^M p_i m_i + \gamma \sum_{i=1} ^M p_i\label{eq10}\ee
Similar to the previous equation (\ref{eq9}) it contains an essential difference in the entropy factors {$S_i$}. These coefficients are computed directly from the time series data for each of the $M$ assets in the portfolio. Thus for each $i$ we compute the Shannon (Boltzmann) entropy associated with the time series:
\be S_i = -\int P_i \lb x \rb \ln P_i \lb x \rb dx \ee
$P_i(x)$ is the distribution function (PDF) associated with the returns of the time series associated with the $i$-th asset. The use of additive entropies is a somewhat simplified assumption. Strictly speaking we should use conditional entropies that include dependencies among the assets. However our approach has the merit that it yields formulas that can be solved analytically. We also believe that the dependencies are, as for the correlation matrix, much smaller than the diagonal elements.
\par One can easy derive the solutions for weights from equation (\ref{eq10}). Reverting to vector notation, we obtain:
\be p_c = \lb C + \alpha S \mathbf{I} \rb^{-1} \lb \gamma \mathbf{1} - \beta m \rb \label{eq12}\ee
$\mathbf{I}$ is a identity matrix, $\mathbf{1}$ is the identity vector and:
\ber&\beta = \frac{m_c \cdot \mathbf{1}^T \tilde{C}^{-1} \mathbf{1} - m^T \tilde{C}^{-1} \mathbf{1}}{m^T \tilde{C}^{-1} \mathbf{1} \cdot \mathbf{1}^T \tilde{C} ^{-1} m - m^T \tilde{C}^{-1} m \cdot \mathbf{1}^T\tilde{C}^{-1}\mathbf{1}} \nonumber \\
&\tilde{C} = C+\alpha S \mathbf{I} \label{eq13}\\
&\gamma = \frac{1+\beta\cdot\mathbf{1}^T\tilde{C}^{-1} m}{\mathbf{1}^T \tilde{C}^{-1}\mathbf{1}} \nonumber
\eer
We have introduced $m_c$ for the level of expected profit (payoff) of the portfolio. If it is not stated differently, the expected return $m_c$ is set to $20\%$ per annum. The Lagrange multiplier, $\alpha$, should be calculated from some other formula with constrain, such as setting constant expected profit of market portfolio (we will show the result of this solution below).

\par In a previous paper \cite{refObV}, we have developed the concept of the objective function that is proportional to the Shannon entropy and the covariance matrix $C$ as the covariance of 'objective values' for the returns, $C^{ObV}$. The objective values, $w(x)$ is defined in terms of the stationary probability distribution for returns, $P(x)$, viz:
\be \mathbf{P} \lb x \rb = \frac{1}{\mathbf{Z}} e^{-\frac{w \lb x \rb}{d}}\ee
where $\mathbf{Z}$ is a normalization factor. Such functions are familiar to physicists and may be derived by minimizing a 'free energy' functional, $F(w(x))$, subject to constraints on the mean value of the objective function, viz:
\be F=\int dx \mathbf{P} \lb x \rb \left[ \ln \mathbf{P} \lb x \rb + \frac{w \lb x \rb}{d} - \delta \right] \ee
Here $\delta$ is a normalization parameter to $\mathbf{P}(x)$. Normalized covariance matrix $C^{ObV}$  and mean values $\set{m_i^{ObV}}$ are due to the normalization procedure which in simple words makes all $\mean{w_i \lb x \rb }$ equal. Here the notation $i$ shows the dependence of objective value on different asset $i$.
\par We use these ideas in this paper noting that in addition we now need also to change average values for the returns $\set{m_i}$ to the set $\set{m_i^{ObV}}$. The reader should read reference \cite{refObV} for the details of the method.
\par Within this approach, the risk is computed as follows:
\be R = \sum_{i,j = 1}^N C_{ij}^{ObV} p_i p_j + \alpha \sum_{i=1}^N S_i p_i^2 \ee
We now propose a further new quantity that measures the 'quality ratio' (similar to Sharp ratio \cite{refSharp}, but standard deviation is replaced by variance) of the portfolio:

\be QR = \frac{profit}{risk} = \frac{\beta\sum_{i=1}^M m_i^{ObV}}{\sum_{i,j=1}^N C_{ij}^{ObV} p_i p_j + \alpha \sum_{i=1}^N S_i p_i^2} \ee
To complete these calculations requires of course knowing the value of the Lagrange multipliers, $\alpha$ and $\beta$.
\par A route to compute $\alpha$ is by standardizing $QR$ ratio to a market portfolio, $QR_m$. Market weights $p_m$ are associated with the market portfolio (weights comes from capitalizations) and balanced by investors with all available information. One can assume that $QR_m$ takes the maximum value, ie.
\be QR_m = \frac{\beta p_m^T m^{ObV}}{p_m^T C^{ObV} p_m + \alpha p_m^T S p_m} \label{eq16}\ee
Putting $QR_m = 1$ we can solve equation (\ref{eq16})
\be \alpha = \frac{\beta p_m^T m^{ObV} - p_m^T C^{ObV} p_m}{p_m^T S p_m} \ee
Note that the parameter $\beta$ depends on $\alpha$ itself (see Eq.~\ref{eq13}). A self consistent solution is then required.

\par
\begin{figure}
\centering
\includegraphics[width=8cm,angle=0]{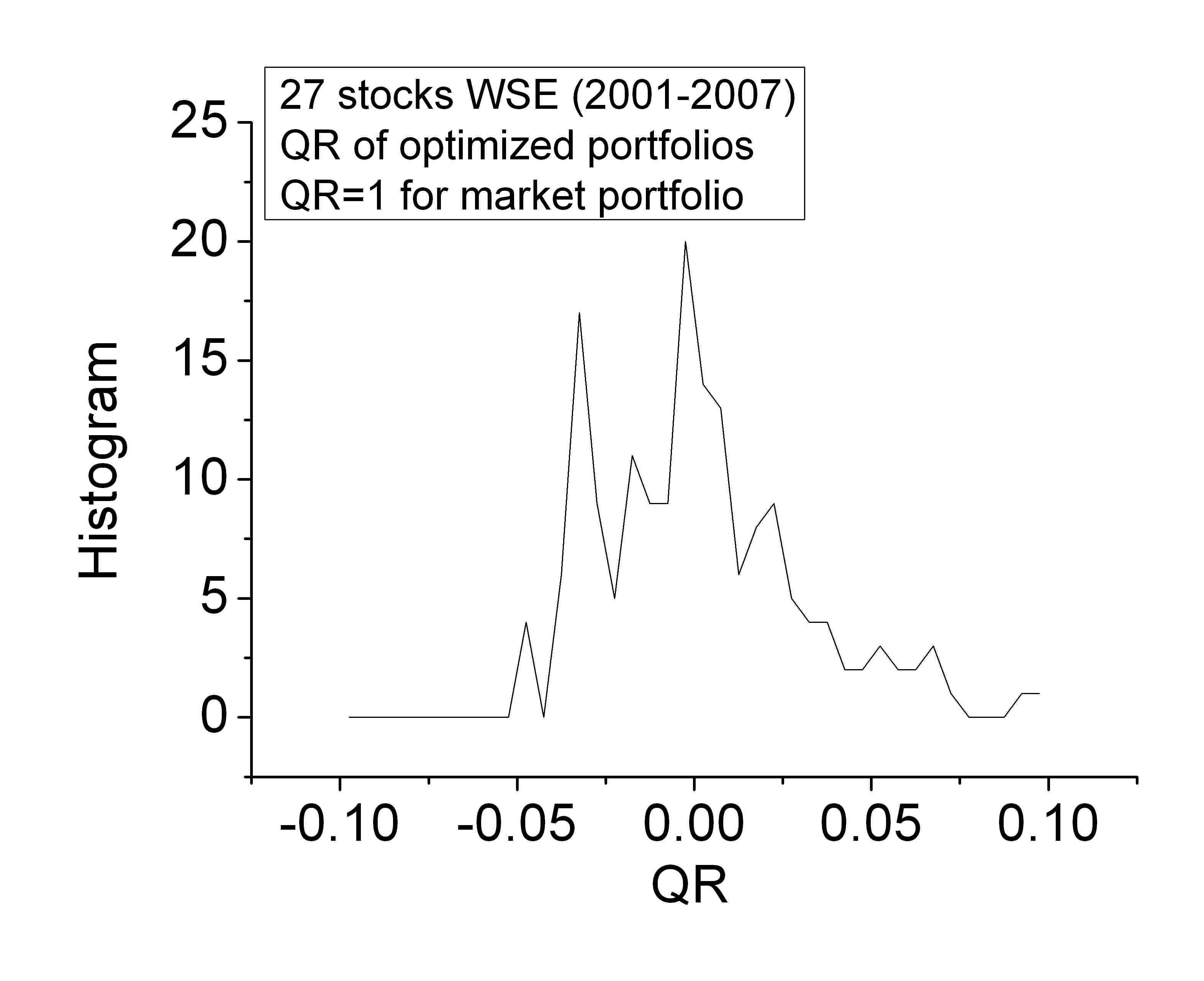}
\caption{The histogram of quality ration $QR$ for optimized portfolio of 27 stocks on Warsaw Stock Exchange in time 2001-2007. The reference value of $QR_m$ is 1 for market portfolio given by values of capitalizations of the firms. \label{fig1} }
\end{figure}

 We present the distribution of $QR$ for the optimized portfolio in Fig.~\ref{fig1} and we can see that $QR$ seems to be small compared to the value of 1 (assumed for the market). This is because the return for the market portfolio over 75 days is much larger (negative or positive) compared with $20\%$ per annum for $m_c$. $QR$ can be negative since over a short period, the profit from market portfolio can also be negative (since market weights are always positive). We stress that the expected profit from the market portfolio $m_c$ is always positive but values calculated from historic returns may differ significantly from the expected return.
\par
\begin{figure}
\centering
\includegraphics[width=8cm,angle=0]{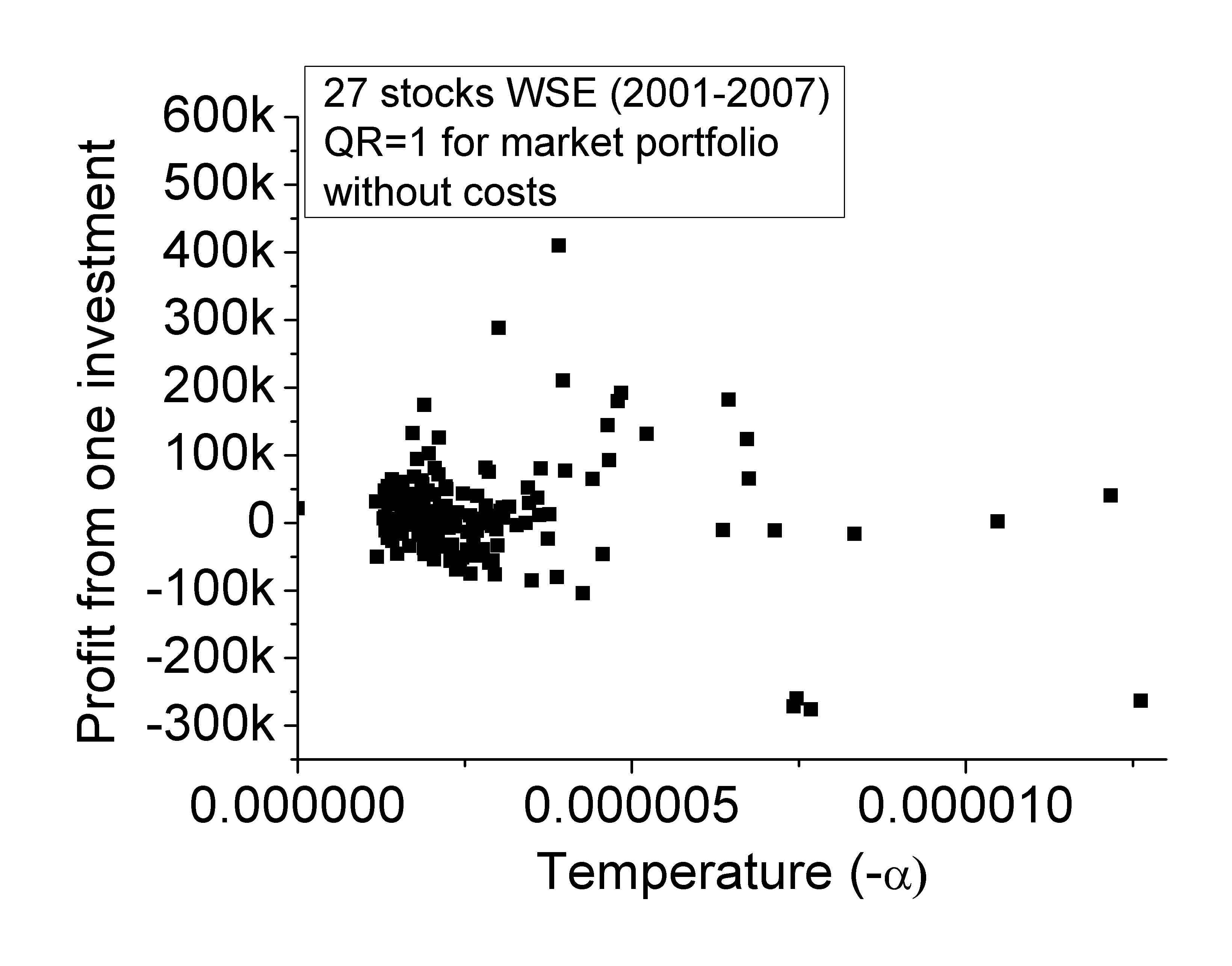}
\caption{The profit values versus temperature of optimized portfolio of 27 stocks on Warsaw Stock Exchange in time 2001-2007. The reference value of $QR_m$ is 1 for market portfolio given by capitalizations of the firms.\label{fig2}}
\end{figure}

\begin{figure}
\centering
\includegraphics[width=8cm,angle=0]{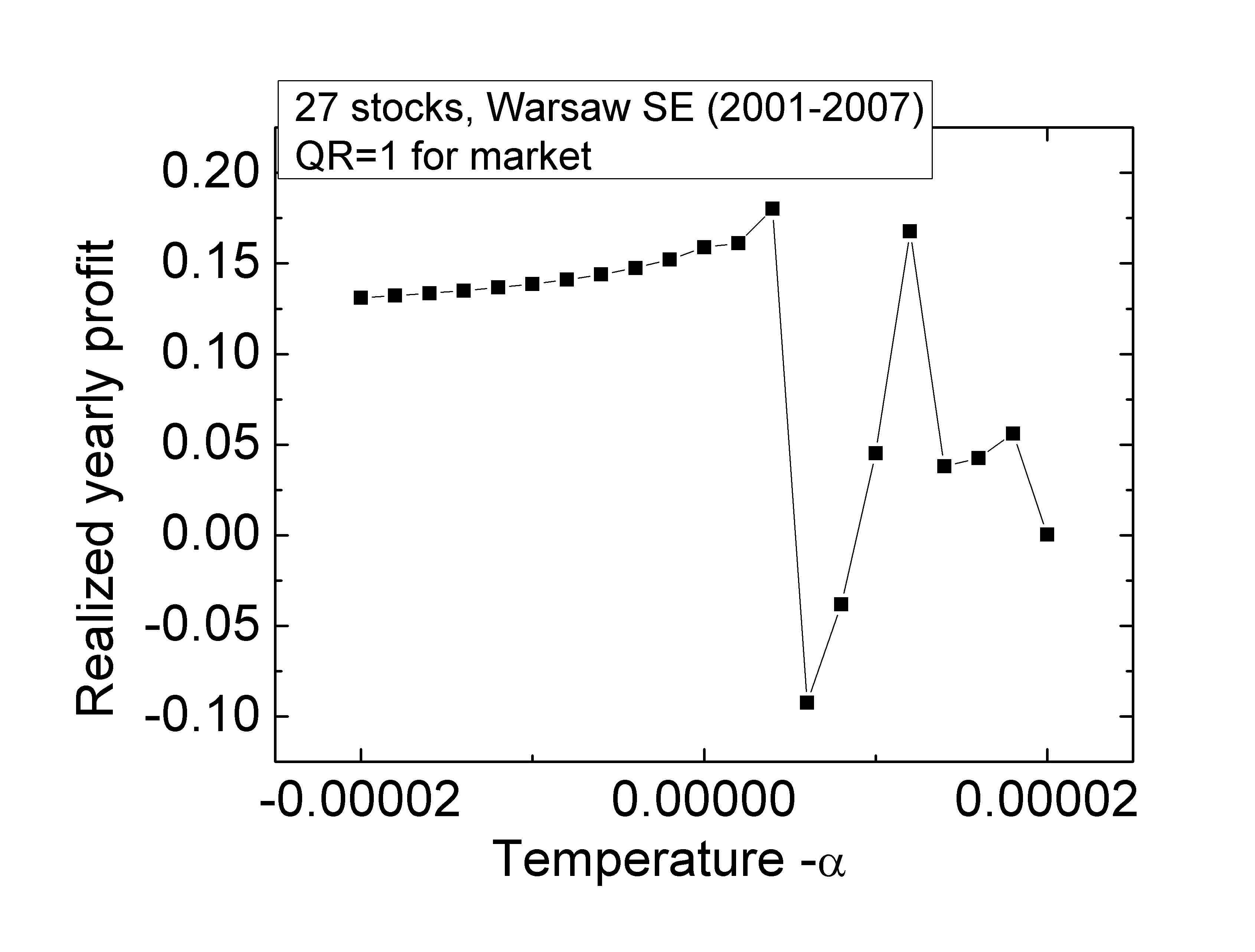}
\caption{The plot of realized yearly profit versus temperature given by Lagrange multiplier to entropy. The profit is given by optimization of portfolio of 27 stocks on Warsaw Stock Exchange in time 2001-2007. The reference value of $QR_m$ is 1 for market portfolio given by capitalizations of the firms.\label{fig3}}
\end{figure}

\begin{figure}
\centering
\includegraphics[width=8cm,angle=0]{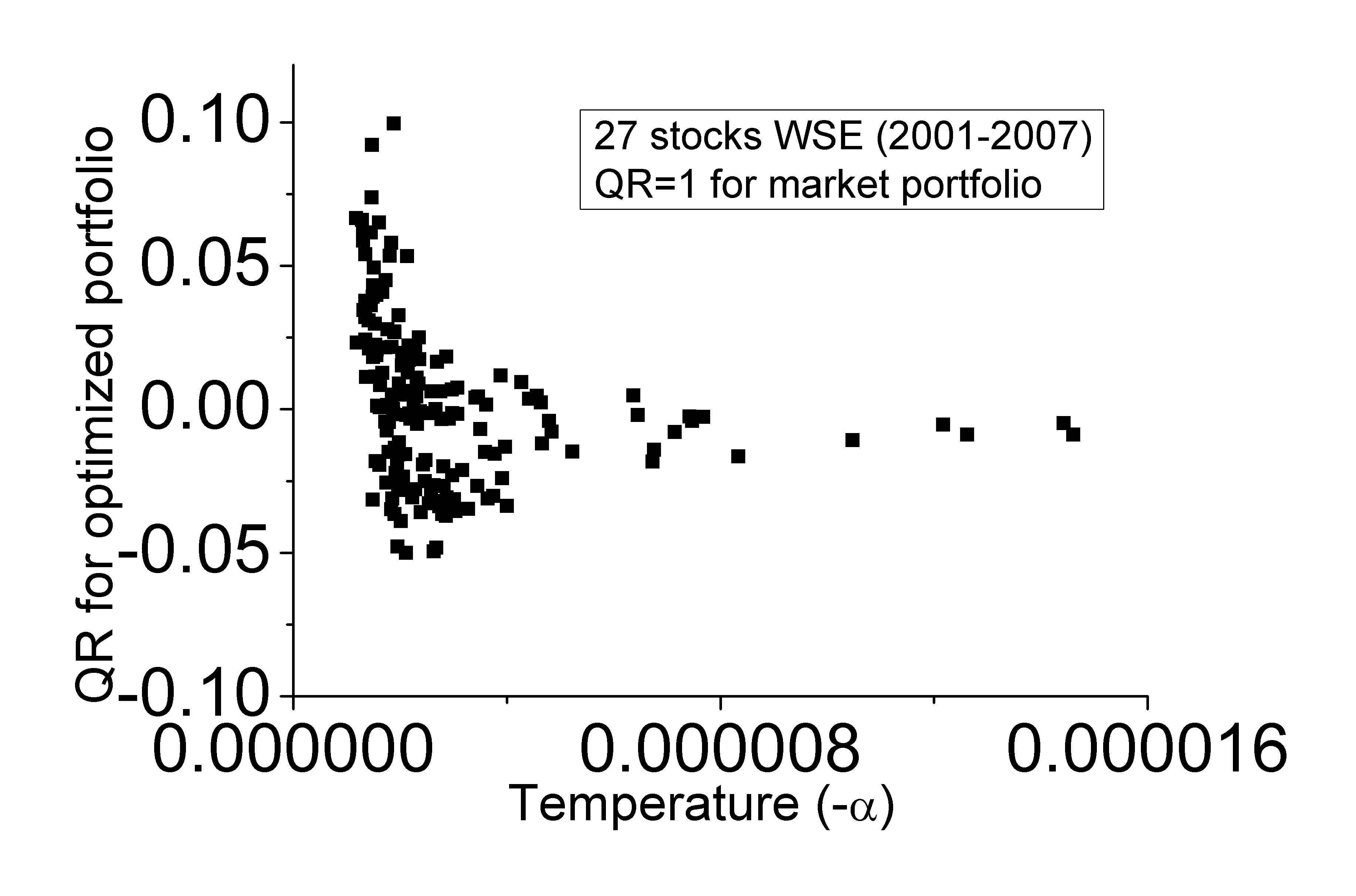}
\caption{The plot of quality ratio $QR$ versus temperature given by Lagrange multiplier to entropy. The optimization is done with 27 stocks on Warsaw Stock Exchange in time 2001-2007. The reference value of $QR_m$ is 1 for market portfolio given by capitalizations of the firms.\label{fig4}}
\end{figure}

 In Fig.~\ref{fig2} we show the dependence of profit from portfolio on the temperature (Lagrange multiplier to entropy) of the market. Here one investment is investment from one day of whole portfolio. We see that risk (the variance of the profit) is an increasing function of temperature. This may be understood in physical terms; a 'hot' market is more volatile and the return more difficult to predict. The same point is seen in Fig.~\ref{fig3} where we show the realized annual profit as a function of the initial set up temperature. For high positive temperatures the profit is much more volatile that for low temperatures, or 'cool' markets, when the portfolio changes much more smoothly. As the temperature increases from small (even negative) values the realized profit increases until at an apparent critical temperature, the profit falls abruptly and behaves in a more volatile manner consistent with high risk as a result of 'overheating'. The situation can be seen in Fig.~\ref{fig4} where we show the $QR$ for the optimized portfolio as a function of the temperature of the market. In this figure we see that high $QR$ ratios are present only in 'cool' markets from which it can be understood that in 'cool' markets the risk is low and the predictability is high.
\par
\begin{figure}
\centering
\includegraphics[width=8cm,angle=0]{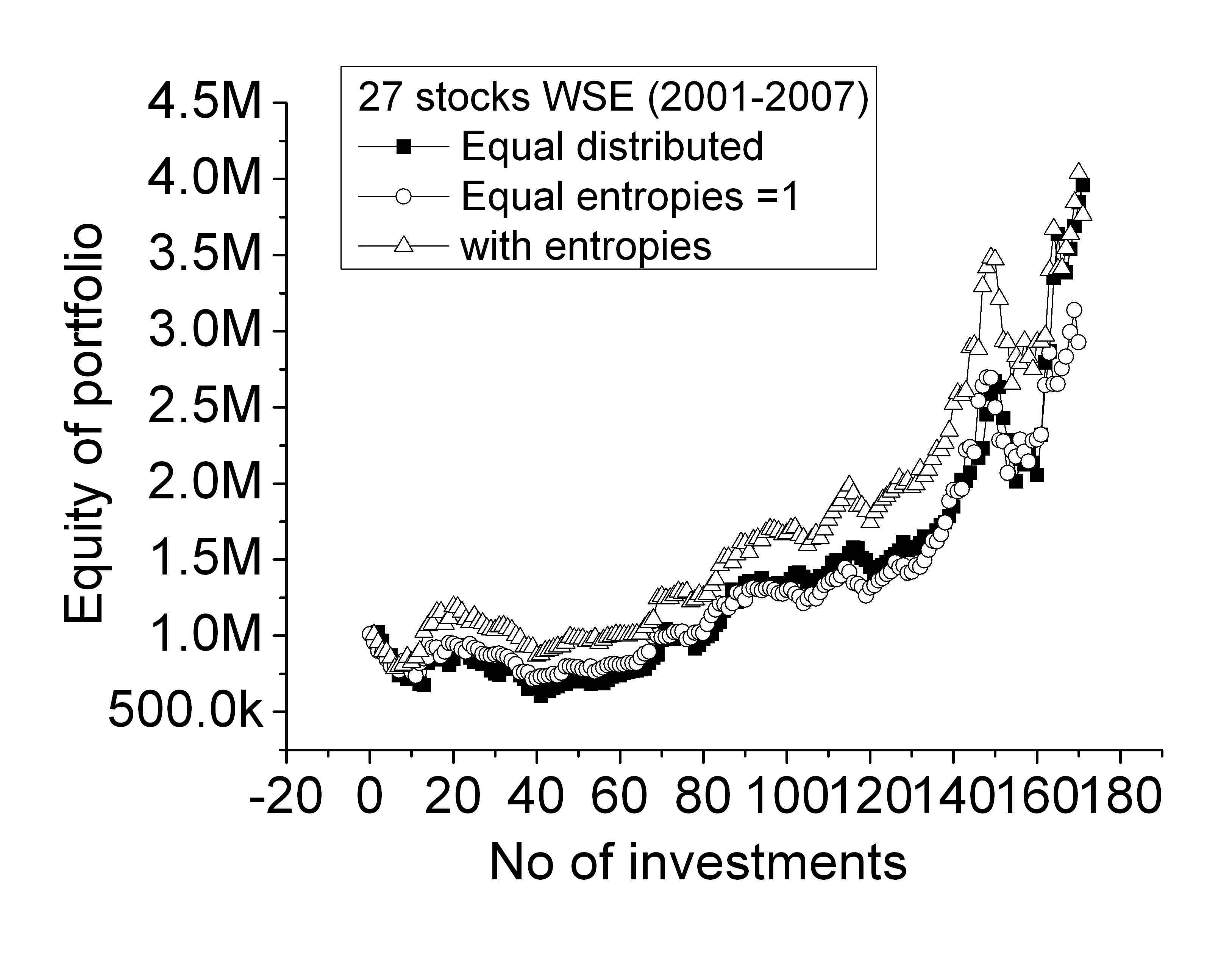}
\caption{The compound value of portfolio given by equal distributed portfolio - solid squares, by optimization procedure (\ref{eq12}) with equal entropies ($S_i=1$) - open circles and by full optimization procedure (\ref{eq12}) - triangles. The optimization is done with 27 stocks on Warsaw Stock Exchange in time 2001-2007. For two last optimizations the reference value of $QR_m$ is 1 for market portfolio given by capitalizations of the firms.\label{fig5}}
\end{figure}

 Fig.~\ref{fig5} shows the runaway of 27 stocks distributed equally within the portfolio. The portfolio optimized using equation (\ref{eq10}) including the entropic corrections gives higher profit than that using equation (\ref{eq9}) that leaves the effect out. We conclude that entropy brings new information about the risk.
\par In the case of equation (\ref{eq10}) there is a shift of accent from average return, $m$, to correlations between the assets. This is caused by negative $\alpha$. When $\alpha$ is negative the diagonal elements of the correlation matrix are less than those for the normal Markowitz approach so the diversification is due to correlations between assets rather than from variances and average returns. The minimization of 'free utility' in the portfolio results on relying on minimization of the risk rather than maximizing the profit (shift the accent from $m$ to correlations what was mentioned). This is because prediction using the return, $m$ fails at high temperature.
\par
\begin{figure}
\centering
\includegraphics[width=8cm,angle=0]{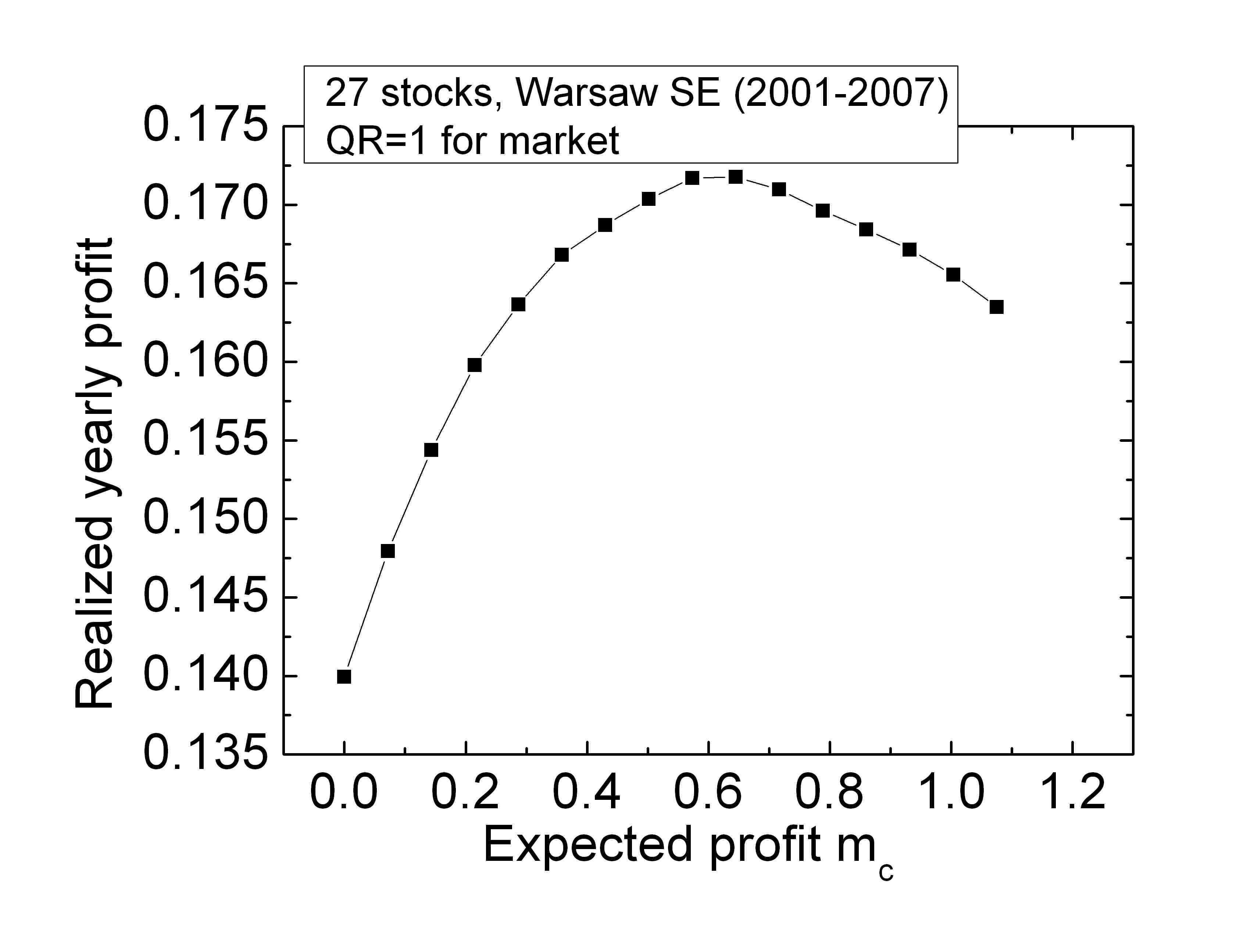}
\caption{\label{fig6} Plot of realized yearly profit with respect to expected profit $m_c$.}
\end{figure}

 Fig.~\ref{fig6} illustrates that the realized profit is a non-monotonic function of the expected profit, $m_c$. The maximum realized profit in this illustration occurs for $m_c~60\%$ which is much larger than the profit of the market portfolio: $20\%$. This suggests that one might aim for large profits bearing in mind that they will be reduced.
\section{Conclusions}
\par
In this Paper we describe the way of entropy inclusion to standard portfolio optimization. The entropy as well as variance reveal the risk value associated to the portfolio past returns. Entropy brings new information into optimization that is increasing its effectiveness. The portfolio constructed with entropy included shows larger profit what is the best proof of properly defined theory.



\newpage

\begin{thebibliography}{99}
\bibitem{refMarkowitz} H Markowitz, \emph{Portfolio Selection, Efficient Diversification of Investments}, Wiley, New York, (1959).
\bibitem{refSiman} Y Simann, \emph{Estimation risk in portfolio selection: The mean variance model versus the mean absolute deviation model} Management Sci. 43 1437/1446 (1997)
\bibitem{refTze} T-L Lai, H Xing and Z Chen, \emph{Mean-Variance portfolio optimization when means and covariances are unknown}, The Annals of Applied Statistics Vol. 5(2A), 798-823 (2011).
\bibitem{refRenyi} A R{\'e}nyi, \emph{On measures of entropy and information}, In Proc. 4th Berkeley Symp. Math. Statist. and Prob.,
Vol. 1, pp. 547-561 (1961).
\bibitem{refShannon} C-E Shannon, \emph{A mathematical theory of communication}, The Bell System Technical Journal Vol. 27, 379-656 (1948).

\bibitem{refBouchaund} J-P Bouchaud and M Potters, \emph{The Theory of Financial Risks; From Statistical Physics to Risk management}, Cambridge University Press, 2000.
\bibitem{refPotter}  J-P Bouchaud, M Potters and J-P Aguilar, \emph{Missing information and Asset allocation}, arXiv:cond-mat/9707042 4 Jul 1997.
\bibitem{refTsallis}  C Tsallis, \emph{Nonadditive entropy: the concept and its use}, European Physical Journal A 40, 257--266 (2009).
\bibitem{refTsallis2} G-A Tsekouras, C Tsallis, \emph{Generalized entropy arising from a distribution of q-indices}, Physical Review E 71, 046144 (2005).
\bibitem{refObV}  K Urbanowicz, P Richmond and J-A Holyst, \emph{Risk evaluation with enhanced covariance matrix},  Physica A 384(2), 468-474 (2007).
\bibitem{refSharp} W-F Sharpe, \emph{Capital asset prices: A theory of market equilibrium under conditions of risk} J. Finance 19, 425/442 (1964).
\end{thebibliography}

\end{document}